# Kinetics of non-equilibrium quasiparticle tunneling in superconducting charge qubits

M. D. Shaw,[1] R. M. Lutchyn,[2] P. Delsing,[3,4] and P. M. Echternach[1,4]

[1] Department of Physics and Astronomy, University of Southern California, Los Angeles, CA, 90089-0484

[2] Joint Quantum Institute, Department of Physics, University of Maryland, College Park, MD, 20742

[3] Microtechnology and Nanoscience, MC2, Chalmers University of Technology, 412 96 Göteborg, Sweden

[4] Jet Propulsion Laboratory, California Institute of Technology, Pasadena, CA, 91109

We directly observe low-temperature non-equilibrium quasiparticle tunneling in a pair of charge qubits based on the single Cooper-pair box. We measure even- and odd-state dwell time distributions as a function of temperature, and interpret these results using a kinetic theory. While the even-state lifetime is exponentially distributed, the odd-state distribution is more heavily weighted to short times, implying that odd-to-even tunnel events are not described by a homogenous Poisson process. The mean odd-state dwell time increases sharply at low temperature, which is consistent with quasiparticles tunneling out of the island before reaching thermal equilibrium.

# I. Introduction

In recent years, mesoscopic single-Cooper-pair devices have attracted considerable interest and demonstrated remarkable technological improvement. Examples of such devices include the Cooper-pair box charge qubit (CPB),[1] the single-Cooper-pair transistor (SCPT),[2,3] and the Cooper-pair pump current standard.[4] In all of these devices, performance is sharply degraded by the undesired tunneling of quasiparticles. In particular, the use of the CPB as a building block for quantum computation requires the preparation and manipulation of delicate superpositions of charge states, which are disrupted when a quasiparticle tunnels across the barrier. While quasiparticle tunneling is not currently the limiting mechanism for decoherence in charge qubits, it is a fundamental source of phase relaxation[5] and imposes a sharp limit on the qubit operation time.

Since single-Cooper-pair devices are typically operated at temperatures below 100 mK, the density of quasiparticles in thermal equilibrium is exponentially suppressed.[2] As such, recent attempts to understand the kinetics of quasiparticle tunneling have focused on the behavior of quasiparticles out of chemical equilibrium, and considerable progress in understanding such systems has recently been made both in theory and experiment.[6,7,8,9,10,11,12]

In this experiment, we directly measure quasiparticle tunneling statistics in the time domain for a pair of CPB charge qubits using an established technique.[8] We extract independent dwell time distributions in the even and odd states, and interpret them using a kinetic model of quasiparticle trapping.[11] We then use this model to understand the behavior of the transition rates as a function of temperature.

## II. Experiment

The device used in these experiments consists of a pair of CPB charge qubits weakly coupled with a fixed capacitor. For the purposes of this experiment, the qubits can be treated as two independent, uncoupled devices fabricated on the same chip. A circuit diagram and SEM images of the two qubits, which we denote "left" and "right", are shown in figure 1. Independent readout of both qubits was performed using a multiplexed quantum capacitance measurement with RF reflectometry. As individual quasiparticles tunnel, the capacitance switches stochastically between two values, which are characteristic of odd and even parity in the device. By analyzing the statistics of this fluctuation in the time domain, we can independently measure the rates of odd-to-even and even-to-odd transitions, corrected for the finite bandwidth of the measurement.[13]

The quantum capacitance readout (QCR) is a dispersive measurement of the reactive response of an LC oscillator coupled capacitively to the qubit island.[14,15] The oscillator is tuned to a frequency much lower than the qubit energy level spacing, minimizing measurement backaction and filtering high frequency noise from the cold amplifier and the RF line. The overall capacitance of the oscillator is

$$C_i = C_T + C_C + \frac{C_g C_J}{C_g + C_J} - \frac{C_g^2}{e^2} \frac{\partial^2 E_i}{\partial n_g^2} \quad (1)$$

when the qubit is in the $i$th energy eigenstate. In this equation, $C_T = 450$ fF is the capacitance of the tank circuit, $C_C = 30$ fF is the capacitance of the input coupling capacitor, $C_g = 2$ fF is the RF gate capacitance, $C_J = 2.2$ fF is the junction capacitance, $E_i$ is the $i$th qubit energy eigenvalue, and $n_g = C_{cg} V_g / e$ is the normalized gate charge, where $C_{cg} = 230$ aF is the control gate capacitance. The third term in Eqn. 1 is referred to as the quantum capacitance, and is proportional to the curvature of the qubit energy level. By measuring the phase shift of a reflected RF signal, one can directly extract the quantum capacitance, which in typical experiments is on the order of 1 fF. Since the ground and first excited state have opposite curvature at the degeneracy point, this dispersive technique can be used to measure the state of the qubit directly at its operating point. In this experiment, a multiplexed QCR was used to read out both qubits concurrently. Two parallel lumped-element LC tank circuits with different inductances are capacitively coupled to a single transmission line, which is probed with a two-tone RF signal. The reflected signal is demodulated in a homodyne technique with two analog quadrature mixers, allowing independent monitoring of the two qubits.

The tank circuits and qubit structures are fabricated on the same R-plane $Al_2O_3$ substrate, as shown in figure 1. The left and right tank circuits have center frequencies of 556.42 and 612.38 MHz, respectively, with Q-factors of approximately 3000. The tank circuit capacitors and inductors have design values of 450 fF and 120 nH for the right tank circuit, and 450 fF and 150 nH for the left. The tank circuit capacitance derives primarily from the parasitic capacitor of the inductor coil, although an on-chip interdigitated capacitor is also present. There are also coupling capacitors between the LC circuits and the 50 Ω transmission line, with nominal values of 30 fF. The tank circuits are patterned with photolithography and made from a superconducting Al/Ti/Au trilayer with respective material thicknesses of 900, 200, and 200 Å. In such a trilayer, the critical temperature $T_C$ is strongly dependent on the thickness of the aluminum layer,[16] so a thin-aluminum trilayer will act as a quasiparticle trap. For a 900 Å aluminum layer, the energy gap $\Delta/k \approx 1.5$ K. In previous samples with a similar design, a Al/Ti/Au trilayer with a 300 Å aluminum layer had an approximate gap energy $\Delta/k = 0.88$ K, with much less severe quasiparticle poisoning. This is consistent with recent studies of the effects of normal-metal traps on quasiparticle tunnel rates.[17]

The qubit structures were patterned using electron-beam lithography and fabricated using a conventional shadow-mask aluminum evaporation technique. The qubits themselves consist of small aluminum islands coupled to ground via ultra-small (100 x 100 nm) Al/AlOx/Al tunnel junctions arranged in a loop, or DC-SQUID configuration. The qubits are controlled by two separate gate capacitors and an externally applied magnetic flux. The island thickness is 25 nm, while the lead thickness is 55 nm. Note that the two qubits have SQUID loops of different sizes, so that the applied flux can be tuned quasi-independently. No attempt was made to engineer the island-lead gap profile by oxygen doping the Al films. By fitting traces of the ground state quantum capacitance, we estimate that $E_c/k = 200$ mK in both qubits, while $E_J^{right}/k = 160$ mK and $E_J^{left}/k = 350$ mK. The two qubits are weakly coupled along $\sigma_{zz}$ with a fixed capacitor, as shown in figure 1. In capacitively coupled CPBs, the coupling energy

$$E_m = \frac{e^2 C_m}{C_{\Sigma 1} C_{\Sigma 2} - C_m^2},$$ where $C_m$ is the mutual qubit coupling capacitance and $C_{\Sigma 1,2}$ is the total island capacitance for the left and right qubits, and can be estimated by measuring

the gate voltage dependence of the ground-state quantum capacitance as the system is brought through the mutual degeneracy point. In this sample, no excursion was found in the left qubit as the right qubit was brought through its degeneracy point, for any value of gate voltage in the left qubit. The same was also true for the right qubit. From this, we estimate that the coupling energy $E_m << E_c, E_J$, and to first approximation the two qubits can be treated as uncoupled. For the remainder of this paper, we will discuss the left and right qubits independently, as separate devices mounted on the same chip. The sample was mounted to the mixing chamber of a dilution refrigerator with a base temperature of 18 mK.

To observe quasiparticle tunneling in the time domain, the downconverted RF signal is amplified, filtered, and digitized with an oscilloscope, where individual tunnel events appear as sudden jumps in the phase of the reflected wave, since the tunneling of a single quasiparticle causes the CPB gate charge to shift by 1 e. This technique was first employed by Naaman and Aumentado to measure quasiparticle tunneling rates in an SCPT.[8] The resulting phase shift record takes the form of a random telegraph signal in the time domain, which is filtered with a Schmitt trigger as shown in Figure 2. From this filtered signal, we extract dwell times for both the odd and even states. By fitting a histogram of these dwell times to different functions, one can determine both the odd-to-even and even-to-odd quasiparticle tunneling rates, and determine the nature of temporal correlations between tunnel events.

To optimize the signal quality, the amplifier filter bandwidth was set to 100 kHz, while typical tunnel rates are on the order of 10 kHz. The bandwidth of the tank circuit is approximately 200 kHz. The lowpass filter tends to skew the dwell time distribution toward longer times, but this can be corrected for by applying the procedure described in Ref. 13. The timebase of the oscilloscope was set to 1 μs per point, and data was recorded in 10,000-point "frames" 10 ms in length. To assemble a typical time record, we would sequentially acquire 100 frames with a ~1 s delay between each frame, and concatenate them into a single 1 s time record, dropping the first and last tunnel events within each frame. Because at long times the tunneling process is approximately Poisson, recording the data in this way simply imposes a highpass filter at 100 Hz. Since typical tunnel rates are on the order of 10 kHz, this will not significantly affect the statistics obtained from the dwell time histograms, since count rates approaching 100 Hz are negligible in most cases.

## III. Theory

To explain the temperature dependence of the quasiparticle tunnel rates, we turn to the kinetic theory recently developed by Lutchyn and Glazman.[11] In the following, we assume that the qubit is operated at the degeneracy point, and that $\Delta_I \approx \Delta_L$, the superconducting energy gaps in the island and the lead, are roughly equal.

### A. Distribution of non-equilbrium quasiparticles

Let us assume that an external energy source such as high-frequency electromagnetic radiation has generated a uniform density of non-equilibrium quasiparticles $n_{qp}$ throughout the sample, which are in thermal equilibrium at temperature $T$ when the device is in the even state. While they are assumed to be in thermal equilibrium, they are not necessarily in chemical equilibrium. As discussed previously,[10] these extra quasiparticles will shift the chemical potential of the leads and island by respective amounts $\delta\mu_{L,I}$, which are related to the quasiparticle density

$$n_{qp} = 2D(E_F)\int_{\Delta_{L,I}}^{\infty} dE \frac{E}{\sqrt{E^2 - \Delta_{L,I}^2}} \left( f(E - \delta\mu_{L,I}) - f(E) \right) \quad (2)$$

through the Fermi function $f(E)$. In this equation, $D(E_F)$ is the normal-metal density of states at the Fermi level, and $E/\sqrt{E^2 - \Delta_L^2}$ is the normalized BCS density of states for an s-wave superconductor. To lowest order in temperature,

$$\delta\mu_{L,I} \approx kT \ln\left(1 + \frac{n_{qp}}{N_{L,I}} \exp\left(\frac{\Delta_{L,I}}{kT}\right)\right) \approx \Delta_{L,I} + kT \ln\left(\frac{n_{qp}}{N_{L,I}}\right) \quad (3)$$

where $N_{L,I} = D(E_F)\sqrt{2\pi\Delta_{L,I} kT}$ is the density of quasiparticle states available in the lead (island). When a quasiparticle tunnels from the lead to the island, however, the total number of quasiparticles on the island goes from $n_{qp}\Omega_I$ to $n_{qp}\Omega_I + 1$, and

$$\delta\mu_I^* \approx kT \ln\left(1 + \frac{n_{qp} + \Omega_I^{-1}}{N_I} \exp\left(\frac{\Delta_I}{kT}\right)\right) \quad (4)$$

where $\Omega_I$ is the volume of the island. Such parity effects in the chemical potential have been well understood for some time.[18]

**B. Lead-to-island tunneling**

Consider the behavior of the qubit in the even state at the degeneracy point, $n_g = 1\,e$, as shown in Figure 3. The energy difference between even and odd states is given by[19]

$$\delta E = \tilde{E}_C \left( M_A(1,-\alpha) - M_A(0,-\alpha) \right) - \tilde{\Delta} \approx E_C - \frac{E_J}{2} - \tilde{\Delta} \quad (5)$$

where $\tilde{\Delta} = \Delta_I - \Delta_L$ is the island-lead gap profile, $M_A(r,q)$ is the characteristic Mathieu function, $\alpha = \frac{E_J}{2\tilde{E}_C}$, $\tilde{E}_C = \frac{E_C}{1 + \frac{3hG_N E_C}{32e^2 \Delta_I}}$ is the renormalized charging energy, and $G_N$ is the normal-state tunneling conductance. If $\tilde{\Delta}$ is small or negative, quasiparticles on the lead will tend to become trapped on the island, as indicated in figure 3. Following Ref. 7, the even-to-odd transition rate, which is equal to the lead-to-island tunneling rate, can be found from Fermi's golden rule:

$$\Gamma_{eo} = \frac{G_N}{e^2} \int_{\max(\Delta_I - \delta E, \Delta_L)}^{\infty} dE \, \frac{E(E + \delta E) - \Delta_L \Delta_I}{\sqrt{(E^2 - \Delta_L^2)((E + \delta E)^2 - \Delta_I^2)}} f(E - \delta\mu_L)(1 - f(E + \delta E - \delta\mu_I)) \quad (6)$$

where $f(E - \delta\mu_L)$ is the Fermi function for a quasielectron in the lead at energy $E$ above the Fermi level, and $(1 - f(E + \delta E - \delta\mu_I))$ is the probability of finding an available state in the island at the same energy. For the parameters of our experiment, the cutoff of the Fermi functions are well below the edge of the gap, so $f(E - \delta\mu_L)(1 - f(E + \delta E - \delta\mu_I)) \approx e^{-(E - \delta\mu_L)/kT}$, which formalizes the assumption that the tunnel rates are dominated by the quasiparticle density in the leads, rather than in the island. Note that this expression assumes that while the quasiparticles are out of chemical equilibrium, they are at all times in thermal equilibrium at temperature $T$. The integrand in Eqn. 6 has the usual form of a product of the density of states in the lead and island multiplied by the lead and island occupation functions, but it is important to note that the density of states in this equation differs from the standard form[18,20] for quasiparticle tunneling across a superconducting junction by a factor of $\frac{1}{2}\left(1-\frac{\Delta_L \Delta_I}{E(E+\delta E)}\right)$. This extra factor is the product of BCS coherence factors which arise from destructive interference between the tunneling of electron-like and hole-like quasiparticles. At the degeneracy point, the ground state of the qubit is given in the two-state approximation by $|g\rangle = \frac{1}{\sqrt{2}}(|n\rangle + |n+2\rangle)$, where $|n\rangle$ is the island charge state with $n/2$ Cooper pairs. As such, the transition $\langle n+1|H_T|n\rangle$ must involve adding a quasielectron, while $\langle n+1|H_T|n+2\rangle$ must involve adding a quasihole, where $H_T$ is the single electron tunneling Hamiltonian. Finally, note that Eqn. 6 includes a factor of two for the electron-hole degeneracy, since the integral is identical for negative energies.

The quantity which is actually measured in experiments is the even-state dwell time distribution $N_{even}(t)$, which is the probability per unit time of a lead-to-island tunnel event conditioned on the CPB being in the even state at $t=0$. The odd-state dwell time distribution $N_{odd}(t)$ can be defined in an equivalent way. As demonstrated in Ref. 11, even-to-odd transitions are exponentially distributed,

$$N_{even}(t) = \Gamma_{eo} \exp(-\Gamma_{eo} t) \quad (7)$$

which corresponds to a homogenous Poisson process, for which individual tunnel events are temporally uncorrelated. This is not surprising, since one intuitively expects that quasiparticle tunneling from the lead is a completely incoherent process.

**C. Island-to-lead tunneling**

For the odd-to-even transitions, the situation is more complex. Consider a quasiparticle tunneling from the lead to the island with energy $E \approx \Delta_L + kT$. Once in the island, the quasiparticle will thermalize to the edge of the gap via inelastic phonon scattering in an average time[21]

$$\tau = \tau_0 \left(\frac{kT_C}{\Delta_I}\right)^3 \left[\frac{1}{3}q^3 + \frac{5}{3}q - \frac{\Delta_I}{2\omega}\left(1 + 4\left(\frac{\omega}{\Delta_I}\right)^2\right)\ln\left(\frac{\omega}{\Delta_I} + q\right)\right] \approx \tau_0 \left(\frac{kT_C}{\Delta_I}\right)^3 \left(\frac{\Delta_I}{\delta E}\right)^{7/2} \quad (8)$$

where $\tau_0 \approx 100$ ns is the characteristic electron-phonon scattering time for Al, $T_C$ is the superconducting transition temperature, $\omega = (\Delta_L + kT)/\hbar$, and $q = \sqrt{\left(\frac{\omega}{\Delta_I}\right)^2 - 1}$. For the parameters of our films, $\tau \approx 0.1$-$1$ ms, which is slower than the observed rates of quasiparticle tunneling. Due to the strong dependence of $\tau$ on $\Delta_I$, which we treat as a fit parameter, it is difficult to estimate $\tau$ with great precision. Since the relaxation and tunneling time scales are comparable, we must consider two characteristic rates for island-to-lead tunneling: the tunneling rate $\Gamma_{oe}^{th}$ for quasiparticles which have settled into thermal equilibrium at the gap edge and escape from the island via thermal excitation, and the rate $\Gamma_{oe}^{el}$ for quasiparticles which tunnel elastically before they have the opportunity to relax to the bottom of the well. Naturally, one would expect $\Gamma_{oe}^{el}$ to dominate on time scales short with respect to $\tau$.

By a similar argument used to derive Eq. 6, the phonon-assisted tunneling rate for thermalized quasiparticles is given by

$$\Gamma_{oe}^{th} = \frac{G_N}{e^2} \int_{\max(\Delta_I - \delta E, \Delta_L)}^{\infty} dE \frac{E(E + \delta E) - \Delta_L \Delta_I}{\sqrt{(E^2 - \Delta_L^2)((E + \delta E)^2 - \Delta_I^2)}} (1 - f(E - \delta\mu_L))f(E + \delta E - \delta\mu_I). \quad (9)$$

To estimate the tunneling rate for the unthermalized quasiparticles, we can no longer use the Fermi distribution for occupied quasiparticle energy states in the island. Although on the time scales of interest energy relaxation via phonon scattering is a continuous process, for simplicity let us calculate the tunneling rate assuming that tunneling is perfectly elastic, i.e. the quasiparticle tunnels out of the island with exactly the same energy it tunneled in with. In this approximation, we take the quasiparticle distribution on the island to be a deltafunction at $E = \Delta_L + kT$, the energy of a "typical" quasiparticle on the lead. This gives the simple expression

$$\Gamma_{oe}^{el} = \frac{G_N}{e^2 D(E_F)\Omega_I} \frac{(\Delta_L + kT)(\Delta_L + kT + \delta E) - \Delta_I \Delta_L}{\sqrt{((\Delta_L + kT)^2 - \Delta_L^2)((\Delta_L + kT + \delta E)^2 - \Delta_I^2)}} (1 - f(\Delta_L + kT - \delta\mu_L)). \quad (10)$$

For the parameters of our experiment, $(1 - f(\Delta_L + kT - \delta\mu_L)) \approx 1$, and $\Gamma_{oe}^{el}$ is simply proportional to the density of states (including the coherence factors described above). Due to the square root singularity, this equation predicts a sharp rise in $\Gamma_{oe}^{el}$ as $T \to 0$, which is also observed in our data.

Having discussed the characteristic tunneling rates for both thermalized and unthermalized quasiparticles, we turn to the odd-state dwell-time distribution, which is the quantity actually measured in the experiment. The survival probability $S_{odd}(t)$ for a quasiparticle to remain in the island for a time $t$ after tunneling from the lead is given by

$$S_{odd}(t) = F(t) + \frac{1}{\sqrt{\pi \tau \Gamma_{oe}^{el}}} \exp\left(-\frac{\Gamma_{oe}^{th} t}{\sqrt{\pi \Gamma_{oe}^{el} \tau}}\right) \quad (11)$$

where

$$F(t) = \frac{\tau}{\sqrt{\pi \Gamma_{oe}^{el} g(\Delta_I + \delta E)}} \int_0^\infty dz \frac{g(\Delta_I + \delta E + kTz)(\Gamma_{oe}^{el}(kTz))^2}{1 + \tau \Gamma_{oe}^{el}(kTz)} \exp(-z - t\Gamma_{oe}^{el}(kTz) - t/\tau) \quad (12)$$

is the component of the distribution due to quasiparticles out of thermal equilibrium. In this expression, $g(E) = E/\sqrt{E^2 - \Delta_I^2}$, and $\Gamma_{oe}^{el}(kTz)$ is the golden rule elastic tunneling rate for a quasiparticle in the island at energy $\Delta_I + \delta E + kTz$. Equation (11) is obtained by considering a master equation for $S_{odd}(t)$ including a collision integral, and taking the long-$\tau$ limit for the solution. The dwell time distribution $N_{odd}(t) = \frac{\partial}{\partial t}(1 - S_{odd}(t))$ is the probability density for a quasiparticle to tunnel out of the island at a time $t$ after tunneling in.

At long times, $F(t) \approx 0$ and the odd-state dwell-time distribution $N_{odd}(t)$ is approximately an exponential. For $t > \tau$, $N_{odd}(t)$ is dominated by thermalized quasiparticles, which are excited out of the well by phonon-assisted tunneling. Since each phonon absorption event is independent and uncorrelated, the long-time limit of the dwell-time distribution is approximately an exponential,

$$N_{odd}(t) \approx \frac{\Gamma_{oe}^{th}}{\pi(\Gamma_{oe}^{el})^2} \exp\left(-\frac{\Gamma_{oe}^{th} t}{\sqrt{\pi}\Gamma_{oe}^{el}\tau}\right). \quad (13)$$

For times $t$ which are short compared to $\tau$, $F(t) \gg \frac{1}{\sqrt{\pi}\Gamma_{oe}^{el}} \exp\left(-\frac{\Gamma_{oe}^{th} t}{\sqrt{\pi}\Gamma_{oe}^{el}\tau}\right)$ and the distribution is dominated by unthermalized quasiparticles. For times longer than a characteristic time

$$t_{ch} = \frac{1}{\Gamma_{oe}^{el}}\left(\frac{\delta E}{2kT}\right)^{2/3}, \quad (14)$$

where the density of states ceases to be a rapidly varying function of energy, the dwell-time distribution is also an exponential, with rate parameter $\Gamma_{oe}^{el}$. This can be seen from

Eqn. 12, since for $t > t_{ch}$, the density of states and hence $\Gamma_{oe}^{el}(E) \approx \Gamma_{oe}^{el}$ is a weak function of energy, so the exponential can be taken outside the integral.

However, for $t < t_{ch}$, the dwell time distribution is not an exponential, but is weighted more heavily toward shorter times. As such, the tunnel events are not Poisson distributed, and are temporally correlated over short times. This occurs because the density of states, and likewise $\Gamma_{oe}^{el}(E)$, is a rapidly varying function of $E$, and cannot be treated as a constant. In the short-time limit, the asymptotic expression for the dwell time distribution is given by

$$N_{odd}(t) \approx \frac{2^{4/3}}{\sqrt{3}} \frac{\Gamma_{oe}^{el}}{(\Gamma_{oe}^{el} t)^{1/3}} \exp\left(-3\left(\frac{\Gamma_{oe}^{el} t}{2}\right)^{2/3}\right). \quad (15)$$

Regardless of its functional form, $N_{odd}(t)$ can be easily related to the mean dwell time

$$\langle t \rangle = \int_0^\infty t N_{odd}(t) dt \bigg/ \int_0^\infty N_{odd}(t) dt = -\int_0^\infty dt \int_0^t dt' N_{odd}(t'). \quad (16)$$

For a more detailed treatment of the theory, the reader is referred to Ref. 11.

Recall that this model assumes a uniform density of non-equilibrium quasiparticles $n_{qp}$, which enter only as a shift in the chemical potential when the system is in thermal equilibrium. In the temperature range of typical quantum computing experiments, the odd-to-even tunneling probabilities are dominated by unequilibrated quasiparticles, and are approximately independent of $n_{qp}$. Meanwhile, the even-to-odd transition rate $\Gamma_{eo} \approx K n_{qp}$ is approximately linear in $n_{qp}$, where

$$K = \frac{G_N}{e^2} \frac{e^{\Delta_L/kT}}{N_L} \int_{\max(\Delta_I - \delta E, \Delta_L)}^\infty dE \frac{E(E+\delta E) - \Delta_L \Delta_I}{\sqrt{((E+\delta E)^2 - \Delta_I^2)(E^2 - \Delta_L^2)}} e^{-E/kT}. \quad (17)$$

## IV. Results

As discussed in section II, we have measured the statistics of telegraph noise due to quasiparticle tunneling in a pair of Cooper-pair boxes at a variety of temperatures, tuned to their degeneracy points. From the quantum capacitance signal, we can extract the even- and odd-state dwell time distributions, an example of which is shown in Figure 4 for two different sample temperatures, 18 and 200 mK. While this data is shown for the right qubit, studies of the left qubit were qualitatively similar. This data agrees well with the theory outlined in section III. At 18 mK, shown in figure 4a, the odd-state distribution clearly deviates from an exponential at short times. The solid green line is a single-parameter least squares fit to Eq. 15, with $\Gamma_{oe}^{el} = 54 \pm 2$ kHz. The vertical red line indicates the characteristic time scale $t_{ch} = 155$ µs, and for $t \ll t_{ch}$ the distribution indeed deviates from an exponential. At 18 mK, the even state distribution is clearly an exponential, as shown in figure 4c, with a single-parameter fit to Eq. 7 yielding $\Gamma_{eo} = 5.7 \pm 0.1$ kHz. Note that since the tunnel rate into the island is an order of magnitude slower than the tunnel rate out of the island, on average the box spends most of its time in the even state. The probability for finding the box in the even state at 18 mK is $P_{ev} = \Gamma_{oe}/(\Gamma_{eo} + \Gamma_{oe}) = 0.90$.

At higher temperature, the non-exponential behavior of the odd-state dwell time distribution is seen only at shorter times, while the mean dwell time distribution extends out to longer times. In figure 4b, the odd-state dwell time distribution is plotted for a sample temperature of 200 mK. At this temperature, $t_{ch} = 63$ µs, as shown with the red vertical line. Observe that the non-exponential component of the dwell time distribution occurs only at very short times, within the first several bins of the histogram. Beyond this point, the distribution is best fit to an exponential, with a characteristic rate $\Gamma_{oe} = 9.3 \pm 0.3$ kHz. This is consistent with the predictions of the theory outlined in section III. The even-state dwell time distribution is fit to Eq. 7 with $\Gamma_{eo} = 9.1 \pm 0.3$ kHz. In this case, since the time scales for tunneling into and out of the box are almost equal, $P_{ev} = 0.51$, and the qubit is "fully poisoned."

To quickly compare the dwell time distributions for different temperatures, we compute the mean dwell time (MDT) $\langle t \rangle$ from the data $N(t)$ using Eqn. 16. This is shown in Fig. 5 for five different mixing chamber temperatures ranging from 18-200 mK. The odd-state mean dwell times are shown as black circles, while the even-state dwell times are shown as red squares. All points except for the MDT at 18 mK have been adjusted to include the finite measurement bandwidth, as discussed in ref. 13. Since this correction scheme assumes that the underlying process is Poisson, it has not been applied to the lowest temperature data point. The correction is typically on the order of 10%. Note that the mean dwell times at 18 and 200 mK agree closely with the values extracted from the theoretical fits in figure 4. At 18 mK, the inverse of the odd state mean dwell time is 54.4 kHz, while for the even state it is 6.2 kHz. At 200 mK, the agreement is a bit coarser, with the inverse of the odd state mean dwell time is 12.9 kHz, and for the even state it is 8.7 kHz. This discrepancy is due to the fit of the odd-state dwell time distribution at 200 mK to an exponential, which ignores the short-time effects.

The lines in figure 5 are plots of the golden rule transition rates discussed in section III. The solid green line is a plot of the elastic odd-to-even tunnel rate $\Gamma_{oe}^{el}$ described in Eq. 10, the dotted blue line is the thermally excited odd-to-even tunnel rate $\Gamma_{oe}^{th}$ described in Eq. 9, and the dashed red line is the even-to-odd tunnel rate described in

Eq. 6. These curves are plotted with free parameters $\Delta_I = 2.5$ K, $\Delta_L = 2.6$ K, and the nonequilibrium quasiparticle density $n_{qp} = 9 \times 10^{18}$ m$^{-3}$, with all other parameters fixed to their nominal values. The qubit parameters $E_C/k = 200$ mK and $E_J/k = 160$ mK are estimated from fits of the quantum capacitance as described in section II. While this data was taken for the right qubit, data taken with the left qubit was qualitatively similar. Furthermore, many data sets were taken at each temperature, as other parameters of the experiment were varied, and the temperature dependence is qualitatively reproducible. Note that to fit the data shown in figure 5, we actually find that the superconducting gap in the island is lower than the gap in the lead, so that the island acts as a trap for quasiparticles.

The data and theory shown in figure 5 show a number of interesting features. While the agreement is not perfect, the theory captures all of the salient features. Most strikingly, the theory predicts a sharp increase in the odd-to-even transition rates at low temperature, which is confirmed in the experiment. This increase arises strictly from the inclusion of unthermalized quasiparticles in the island and does not appear in models which assume instantaneous thermalization to the gap edge.[20] As such, we may conclude that for this particular set of parameters, the quasiparticles in the island are out of thermal equilibrium, while the quasiparticles in the lead are in thermal equilibrium but out of chemical equilibrium. At low temperatures, the probability that a quasiparticle in the island will reach thermal equilibrium before tunneling out again becomes small, and the thermally exited tunnel rate is exponentially suppressed, so the tunneling kinetics are dominated by the elastic tunnel events. Since the elastic tunnel rate described in Eq. 10 is approximately proportional to the density of states, $\Gamma_{oe}^{el}$ displays a square root singularity as $T \to 0$. While the increase in the observed rates at low temperatures is sharper than that observed in $\Gamma_{oe}^{el}$, this is reasonable given that the dwell-time distribution is dominated by the non-exponential component at low temperatures. When the dwell-time distribution is approximately exponential, the tunnel rate and the inverse of the mean dwell time $\langle t \rangle^{-1}$ are approximately equal. For the short-time asymptote of the dwell time distribution described in Eq. 13, $\langle t \rangle^{-1} = \frac{3\Gamma_{oe}^{el}}{\sqrt{\pi}}$. Including this factor, the theory predicts that the mean odd state dwell time at 18 mK is 38 kHz, which is closer to the observed value than $\Gamma_{oe}^{el}$ alone. In this data, no attempt has been made to correct for electron heating due to reduced electron-phonon coupling at the lowest mixing chamber temperatures.

Another interesting feature is that the $\Delta_I \Delta_L$ terms in Eqs. 6, 9, and 10, which arise from destructive interference between electronlike and holelike quasiparticle tunneling, are absolutely necessary to fit the data. Early attempts to fit the data using rates without the BCS coherence factors[20] could not simultaneously predict the even-to-odd and odd-to-even transition rates as a function of temperature. While quasiparticle transition rates are suppressed due to destructive interference when the qubit is in its ground state, the theory predicts that both rates will be enhanced in the excited state, adversely affecting qubit performance.

Also note that at low temperatures, the even-to-odd transition rates remain monotonic with temperature, implying that the quasiparticles on the lead are in thermal equilibrium. As a result of the shift $\delta\mu_L$ in the chemical potential of the leads, $\Gamma_{eo}$ does not go to zero as $T \to 0$. In the absence of a nonequilibrium quasiparticle population on the leads, $n_{qp} \to 0$, and the theory recovers the exponential suppression of quasiparticle tunneling at low temperatures.

At higher temperatures, the theory predicts that both the odd-to-even and even-to-odd transition rates will increase exponentially with temperature, as equilibrium quasiparticle states begin to become thermally occupied. While the data does not extend to high enough temperatures to determine the functional form of this increase, a data set taken at 300 mK shows no telegraph signal at all, presumably because both transition rates are much faster than the bandwidth of the measurement. From a simple thermodynamic argument,[3] we expect this transition to occur at a temperature $T^*_{I,L} = \Delta_{I,L} / \ln(N_{I,L} \Omega_{I,L})$, where the free energy difference between even and odd states goes to zero. Since the lead and the island have different volumes, we would expect that this transition occurs at different temperatures for the odd-to-even and even-to-odd transition rates. Based on our estimates for the effective island and lead volumes $\Omega_I \approx 0.5 \mu m^3$ and $\Omega_I \approx 100 \mu m^3$, we obtain $T^*_I = 170\,\text{mK}$ for the odd-to-even transitions, and $T^*_L = 130\,\text{mK}$ for the even-to-odd transitions. However, from the data shown in Figure 5 the exponential increase in the rates does not occur below 200 mK.

The kinetic theory of non-equilibrium quasiparticle tunneling described above can also be used to understand the severe quasiparticle poisoning observed in the differential single cooper-pair box (DSCB).[22] The DSCB is an isolated structure consisting of two small qubit islands separated by a pair of tunnel junctions in a loop configuration, so that the relevant quantum states are the differential charge states between the two islands. Since the entire structure is isolated from ground, it was initially believed that quasiparticle tunneling would be partially suppressed, since there are no "leads" for quasiparticles to tunnel in from. However, non-equilibrium quasiparticles generated on the island themselves are free to tunnel back and forth elastically between the two islands, spending the majority of their time out of thermal equilibrium. In fact, out of four DSCB devices tested, all showed e-periodic staircases characteristic of quasiparticle poisoning. A differential layout of this type is also employed in the optimized "transmon" qubits, where quasiparticle tunneling is also observed at low temperatures.[23]

## V. Conclusions

Using the method of Naaman and Aumentado,[8] we have measured quasiparticle tunneling rates in the time domain for a pair of single Cooper-pair boxes fabricated on the same chip. In studies of the dwell time distribution as a function of temperature near the even-state degeneracy point, we have experimentally verified the theory of quasiparticle tunneling developed by Lutchyn and Glazman.[11] This model gives sound physical insights into the kinetics of quasiparticle trapping and tunneling, and resolves some of the apparent mysteries of low-temperature quasiparticles in single Cooper-pair devices. In particular, we observe a non-Poissonian odd-state dwell time distribution and an increase in the odd-to-even transition rates at low temperature. From this analysis, we conclude that at low temperature, a quasiparticle on the island may be out of thermal as well as chemical equilibrium, while quasiparticles on the leads are in thermal equilibrium at the even-state degeneracy point.

Quasiparticle tunneling at low temperatures is a major problem for the performance of single Cooper-pair devices, and understanding the fundamental physics of non-equilibrium tunneling processes is essential to effective qubit design. Several techniques have been investigated to reduce the rates of quasiparticle tunneling, such as engineering the superconducting gap profile between the lead and the island via oxygen doping[6] and control of film thickness.[24] Another approach is the use of quasiparticle traps[17] and SIN cooling junctions to reduce the population of nonequilibrium quasiparticles on the leads. Still another approach is to more carefully isolate the sample from electromagnetic noise and radiation, which has recently been shown to have a strong effect on quasiparticle tunneling.[25,26]

## VI. Acknowledgements

We are grateful to Ben Palmer for many helpful discussions and a critical reading of the manuscript. We would also like to thank Gerd Schön, Jonas Zmuidzinas, Göran Johansson, Juan Bueno Lopez, and Michael Mandelberg for their helpful comments. We are also grateful for the invaluable help of Justin Schneiderman in the early stages of this research. We would like to thank Richard Muller for performing the electron-beam lithography. This work was conducted at the Jet Propulsion Laboratory, California Institute of Technology, under a contract with the National Aeronautics and Space Administration, and was funded by a grant from the National Security Agency. Roman Lutchyn is supported by a fellowship from the Joint Quantum Institute. Copyright 2008, California Institute of Technology. Government sponsorship acknowledged.

# Figures

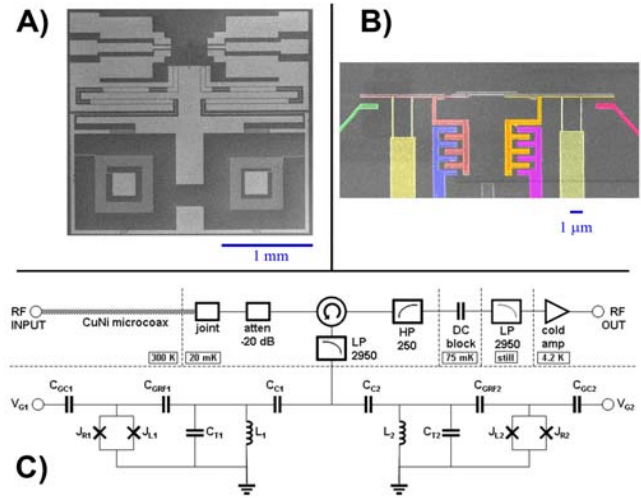

**Figure 1**. (color online) **A)** Scanning electron micrograph of multiplexed on-chip LC oscillators in a similar device to that used in this experiment. The qubit features are at the center. **B)** Scanning electron micrograph of qubit structures in a similar device to that used in this experiment. *Red:* Left qubit island. *Blue:* Left RF gate. *Green:* Left control gate. *Yellow:* Qubit leads and ground plane. *Orange:* Right qubit island. *Pink:* Right RF gate. *Dark Blue:* Right control gate. **C)** Schematic diagram of qubit and quantum capacitance readout circuitry. Nominal component values are given in the text.

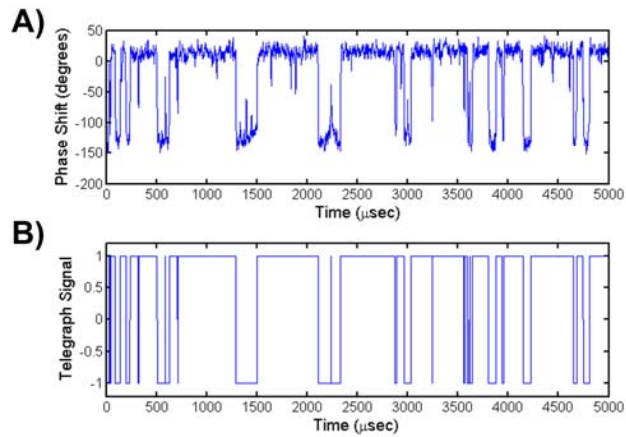

**Figure 2**. (color online) A representative quantum capacitance data trace and the filtered telegraph signal. Dwell time records are assembled by counting the time between zero crossings.

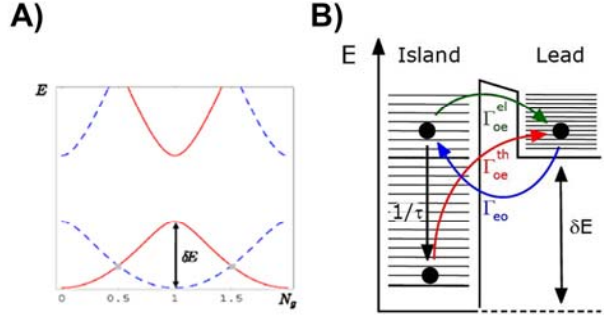

**Figure 3**. (color online) **A)** Energy diagram for the even and odd states as a function of gate voltage. Solid red curves are the first and second even state energy levels, and blue dashed curves are the odd state levels. The odd state levels are simply the even state levels shifted in gate charge by one electron. In the presence of a island-lead gap difference, the blue curve will also be shifted vertically by an energy $\widetilde{\Delta}$. Adapted from Ref. 11.

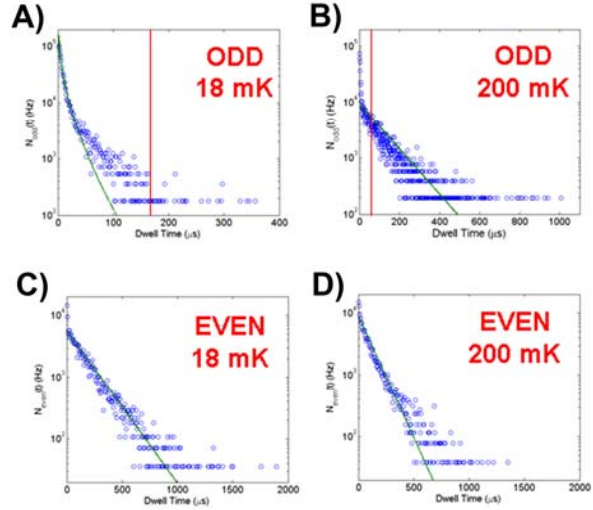

**Figure 4.** (color online) Odd and even state dwell time distributions, measured experimentally for two different mixing chamber temperatures. **A)** Odd state distribution, 18 mK. Fit to Eq. 15 with $\Gamma_{oe}^{el} = 54 \pm 2$ kHz. Vertical red line indicates $t_{ch} = 155$ μs. **B)** Odd state distribution, 200 mK. Fit to exponential with $\Gamma_{oe} = 9.3 \pm 0.3$ kHz. Vertical red line indicates $t_{ch} = 63$ μs. **C)** Even state distribution, 18 mK. Fit to Eq. 7 with $\Gamma_{eo} = 5.7 \pm 0.1$ kHz. **D)** Even state distribution, 200 mK. Fit to Eq. 7 with $\Gamma_{eo} = 9.1 \pm 0.3$ kHz. See section IV for details.

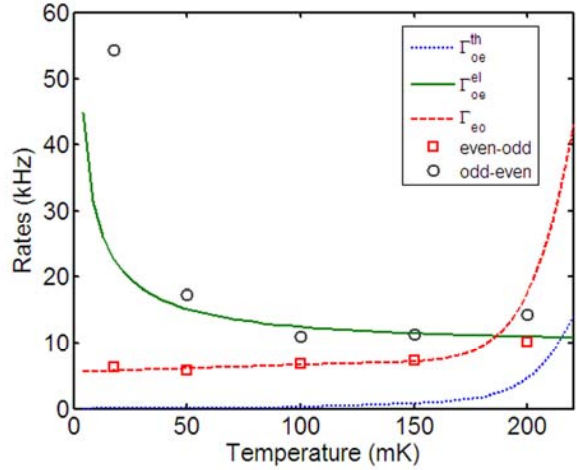

**Figure 5.** (color online) *Points:* Inverse mean dwell times extracted from dwell time distributions as a function of temperature. Black circles are odd-to-even transition rates, red squares are even-to-odd rates. *Curves:* Theoretical estimates of the underlying physical tunneling rates, with the parameters $\Delta_I = 2.5$ K, $\Delta_L = 2.6$ K, and $n_{qp} = 9 \times 10^{18}$ m$^{-3}$. Solid green curve is a plot of Eq. 10, dotted blue curve is a plot of Eq. 9, and dashed red curve is a plot of Eq. 6. Note the negligible contribution of $\Gamma_{oe}^{th}$ at low temperature. Details can be found in section IV.